\documentclass[conference]{IEEEtran}
\usepackage{times}
\usepackage{soul}
\usepackage{helvet}
\usepackage{courier}
\usepackage{array}
\usepackage{booktabs} 
\usepackage{multirow}
\usepackage[english]{babel}
\usepackage[colorlinks,linkcolor=blue]{hyperref}
\usepackage[utf8]{inputenc}
\usepackage[linesnumbered,ruled,vlined]{algorithm2e}
\usepackage{algpseudocode}
\usepackage{subfigure}
\usepackage{graphicx}
\usepackage{amsmath}
\usepackage{bm}
\usepackage{url}
\usepackage{stmaryrd}
\usepackage{balance}
\usepackage{amssymb}
\usepackage{pgfplots}
\usepackage{pifont}
\usepackage{epsfig}
\usepackage{booktabs}
\usepackage{pgffor}
\usepackage{tikz}
\usepackage{multirow}
\usepackage{amsmath}

\newtheorem{defn}{\textsc{Definition}}

\usepackage{tikz}
\usetikzlibrary{positioning,arrows,calc}
\usepackage[all]{xy}
\usepackage{makecell}
\usepackage{enumitem}
\usepackage{color}
\usepackage{array}
\usepackage{booktabs}

\newcommand{\our}{\textsc{HGAT}}

\newcommand{\concatenate}{\operatornamewithlimits{\|}}

\begin{document}
	

\title{Fake News Detection on News-Oriented Heterogeneous Information Networks through Hierarchical Graph Attention}
\author{\IEEEauthorblockN{Yuxiang Ren}
	\IEEEauthorblockA{\textit{IFM Lab} \\
		\textit{Department of Computer Science}\\
		\textit{Florida State University}\\
		Tallahassee, FL, USA \\
		yuxiang@ifmlab.org}
	\and
	\IEEEauthorblockN{Jiawei Zhang}
	\IEEEauthorblockA{\textit{IFM Lab} \\
		\textit{Department of Computer Science}\\
		\textit{Florida State University}\\
		Tallahassee, FL, USA \\
		jiawei@ifmlab.org}
}

\maketitle
	
\begin{abstract}
The viral spread of fake news has caused great social harm, making fake news detection an urgent task. Current fake news detection methods rely heavily on text information by learning the extracted news content or writing style of internal knowledge. However, deliberate rumors can mask writing style, bypassing language models and invalidating simple text-based models. In fact, news articles and other related components (such as news creators and news topics) can be modeled as a heterogeneous information network (HIN for short). In this paper, we propose a novel fake news detection framework, namely \textbf{H}ierarchical \textbf{G}raph \textbf{A}ttention \textbf{N}etwork(\textbf {{\our}}), which uses a novel hierarchical attention mechanism to perform node representation learning in HIN, and then detects fake news by classifying news article nodes. Experiments on two real-world fake news datasets show that {\our} can outperform text-based models and other network-based models. In addition, the experiment proved the expandability and generalizability of {\ our} for graph representation learning and other node classification related applications in heterogeneous graphs.
\end{abstract}

\section{Introduction}\label{sec:introduction}
With explosive growth, fake news has already caused severe threats to the public's factual judgment and governments' credibility. Especially with the wide use of social platforms, they facilitate the generation and dissemination of fake news. For example, during the 2016 US presidential election, numerous fake news about presidential candidates is spread on various social platforms \cite{JCGZWL17}. For example, 115 pro-Trump fake stories shared on Facebook a total of 30 million times, and 41 pro-Clinton fake stories shared a total of 7.6 million times are observed in \cite{AG17}. Such a massive amount of widely spread fake news has greatly destroyed candidates' public persona and misled voters' judgment. It has become very critical to detect fake news on social media in time to block the spread.

\begin{figure}[t]
	\centering
	\begin{minipage}[l]{1\columnwidth}
		\centering
		\includegraphics[width=\textwidth]{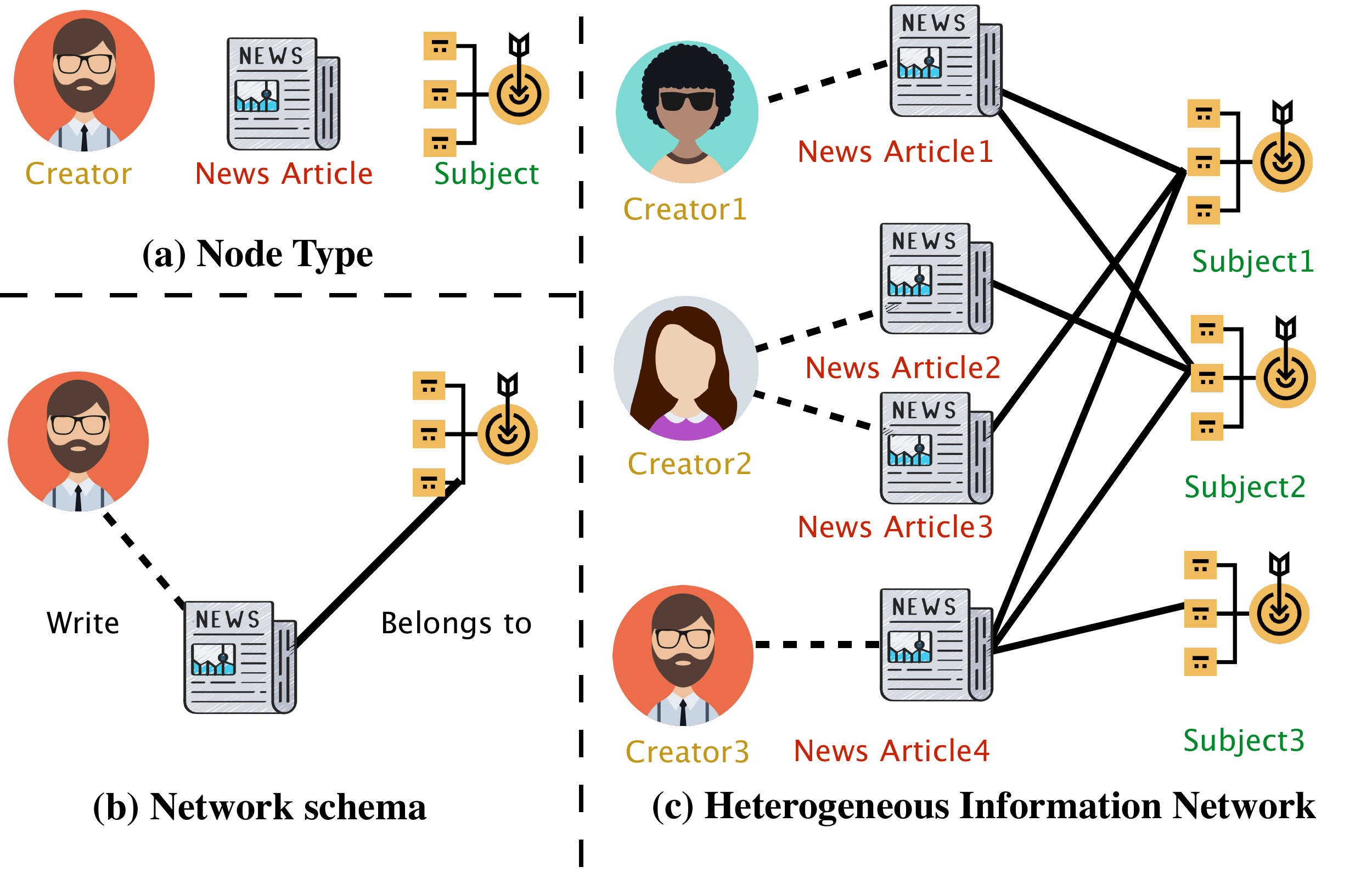}
	\end{minipage}
	\vspace{-15pt}
	\caption{An illustrative example of a heterogenous information network based on PoliticFact data (News-HIN). (a) Three types of nodes (i.e., Creator, News article, Subject). (b) Network schema of News-HIN (c) A  News-HIN consists three types of nodes and two types of links. 
	}\label{fig:example}
	\vspace{-20pt}
\end{figure}
There are significant differences between fake news and traditional fraudulent information. First, fake news is intentionally edited by the creators to achieve the purpose of misleading the public. For instance, news about the same event published by different creators is highly similar in most content, but fake news carries malicious content in objective statements. Although the proportion of these malicious contents is negligible, this is enough to make the news a harmful fake one. Second, for traditional suspicious information, like spam \cite{XWLP12_KDD}, people instinctively have a precautionary mentality that makes themselves less likely to be deceived. But for news, people usually actively search, receive, and share without being on guard about authenticity. Third, spams usually are easier to be detected because of the abundant regular messages; yet, detecting fake news is incredibly challenging since news is very time-sensitive. The evidence collected about past news can not benefit the detection of emerging fake news apparently.  

These characteristics of fake news make the detection more challenging. In order to detect fake news more effectively, it is necessary to mine meaningful information from different views instead of focusing on the news contents solely. In fact, fake news does not exist independently in the form of articles. For example, news creators and news subjects also exist in online social media. The information from news creators and news subjects describes the news in a more comprehensive view and helps us more thoroughly eliminate fake news and relating components. In detail, for the news creators, we can collect profile information and other supplementary knowledge. As for the news subjects, background knowledge and other related information can be collected to support fake news detection. News articles and other related components can be modeled as a heterogeneous information network (HIN for short) \cite{SLZSY17}. HINs have a powerful capability of representing rich information, and we formulate fake news detection as the node classification problem in the HIN in this paper. We present an illustrative example of a news-oriented heterogeneous information network (News-HIN) in Figure~\ref{fig:example}.

\noindent \textbf{Problem Studied}: In this paper, we propose to study the HIN-based fake news detection problem. 
We model the fake news detection problem as a node classification task in the HIN, which requires us to learn the more comprehensive and discriminative representation of news article nodes.

The main challenges of the fake news detection problem in the HIN lie in the following three points:  

\begin{itemize}
	\item \textit{Heterogeneity}: There exist various types of heterogeneous information related to news articles. Learning effective node representations in a HIN in a unified way is not an easy task.
	\item \textit{Hierarchy}: Representation learning in heterogeneous graphs will be a multi-level work because node contents and the information of the schema are contained at different levels. 
	\item \textit{Generalizability}: To ensure the proposed model's applicability to different types of HINs, we need to propose a general learning model that can be extensible to various learning settings. 
	\vspace{-5pt}
\end{itemize}

To handle these challenges aforementioned, we propose a novel \textbf{H}ierarchical \textbf{G}raph \textbf{A}ttention \textbf{N}etwork ({\our}) to detect fake news. {\our} employs a hierarchical attention mechanism to learn the representation of news article nodes. Based on the learned node representation, fake news can be identified through the node classification task. In particular, for each news article node, we use the node-level attention mechanism to learn a set of weights for its neighbors of the same type. Using these sets of weights, we aggregate neighbor nodes of the same type into a schema node. The schema-level attention works to learn the attention weights of different schema nodes. Based on the two-level attention, {\our} can get the optimal combination of different types of neighbors in a hierarchical manner. The learned node representations capture the features from different heterogeneous information sources. {\our} can be optimized in an end-to-end manner by backpropagation. 

The contributions of our work are summarized as follows:
\begin{itemize}
	\vspace{-4pt}
	\item We attempt to detect the fake news in the heterogeneous information network with the support of the heterogeneous graph neural network, while without handcrafted features (e.g., meta-path).
	\item We propose the novel {\our} model, which takes different types of node contents and diverse categories of connections into consideration simultaneously. {\our}, as a general model for representation learning, has excellent potential to be applied to other applications in the HIN. 
	\item We conduct extensive experiments on two real-world datasets to demonstrate the effectiveness of {\our}. 
\end{itemize}

\section{Related Work} \label{sec:related_work}

\subsection{Fake News Detection}
As an emerging topic, some research works have been proposed. Among them, the knowledge-based approach aims to assess the authenticity of news by comparing the knowledge extracted from the news contents with real knowledge \cite{GPLJFA15}. Yet, the timeliness and integrity of the knowledge map remain an unresolved issue \cite{ZZ18}. Another typical way is based on writing style, such as discourse level by employing rhetorical structure theory \cite{VRTL15}, and sentiment \& readability \cite{VBAR17}. Based on relationships among news articles, users (spreaders) and user posts, matrix factorization \cite{SWL19}, tensor factorization \cite{SRMS18}, hierarchical word encoder \cite{cui2019defend}, and Recurrent Neural Networks (RNNs) \cite{ZCFG18} have been developed for fake news detection. 

\subsection{GNNs and Network Embedding}
Graph Neural Networks (GNNs) for representation learning of graphs learn nodes' effective feature vectors through a recursive neighborhood aggregation scheme \cite{XHLJ10}. Kipf et al. \cite{TM17} propose Graph Convolutional Network (GCN). Graph Attention Network (GAT) \cite{VCCRLB18} first imports the attention mechanism into graphs. However, the above graph neural networks are presented for the homogeneous graphs. Wang et al. \cite{WJSWCYY19} consider the attention mechanism in heterogeneous graph learning through the model HAN. However, meta-path as a handcrafted feature limits HAN, and HAN ignores node contents carried by other types of nodes. The learned embeddings from network embedding methods can be applied to the downstream tasks \cite{WJSWCYY19}. Some models have been proposed to deal with homogeneous networks, including the random walk based methods \cite{WPMDJ18}, the matrix factorization based methods \cite{XPJJWS17}, the deep learning-based methods \cite{DPW16}. In order to handle the heterogeneity, metapath2vec \cite{YNA17} samples random walks under the guidance of meta-path on heterogeneous graphs. 
\vspace{-5pt}

\section{Concept and Problem Definition} \label{sec:formulation}
\vspace{-5pt}
In this section, we use the \textit{PolitiFact} data as an example to introduce some terminologies used in this paper. 
These concepts are equally applicable to other datasets.
\vspace{-5pt}
\subsection{Terminology Definition}
\textit {PolitiFact} dataset contains three types of entities: news articles, subjects, and creators. They can be modeled as three types of nodes in a heterogeneous network, and different types of links are constructed according to the connections between them. 
\begin{defn}
	(News Articles): News articles refer to the news contents post on social media or public platforms. News articles can be represented as set $\mathcal{N} = \{n_1, n_2,\cdots, n_m\}$. For each news article $n_i \in \mathcal{N}$, it contains textual contents.
\end{defn}

%
\begin{defn}
	(Subject): Subjects usually denote news articles' central ideas, which are the main objectives of writing news articles. The set of subjects can be denoted as $\mathcal{S} = \{s_1, s_2,\cdots, s_k\}$. For each subject $s_i \in \mathcal{S}$, it contains the textual description.
\end{defn}
%
\begin{defn}
	(Creator): Creators denote users who write news articles. The set of creators can be represented as $\mathcal{C} = \{c_1, c_2,\cdots, c_n\}$. For each creator $c_i \in \mathcal{C}$, it contains the profile information, including titles, political party membership, and geographical residential locations.
\end{defn}
A formal definition of News Oriented Heterogeneous Information Networks can be proposed as follows:
\begin{defn}
	(News Oriented Heterogeneous Information Networks (News-HIN)): The news oriented heterogeneous information network (News-HIN) can be defined as $\mathcal{G} = (\mathcal{V}, \mathcal{E})$, where the node set $\mathcal{V} = \mathcal{C} \cup \mathcal{N} \cup  \mathcal{S}$, and the link set $\mathcal{E} = \mathcal{E}_{c,n} \cup \mathcal{E}_{n,s}$ involves the "Write" links between creators and news articles, and the "Belongs to" links between news articles and subjects.
\end{defn}
In order to better understand the News-HIN and utilize type information, it is necessary to define the schema-level description. 
\begin{defn}\label{def:schema}(News-HIN Schema): 
	Formally, the schema of the given News-HIN $\mathcal{G} = (\mathcal{V}, \mathcal{E})$ can be represented as $S_{\mathcal{G}} = (\mathcal{V}_{type}, \mathcal{E}_{type})$, where $\mathcal{V}_{type}$ and $\mathcal{E}_{type}$ denote the set of node types and link types respectively. Here, $\mathcal{V}_{type} = \{\phi_n,\phi_c,\phi_s\}$ and $\mathcal{E}_{type}= \{\textit{Write}, \textit{Belongs to}\}$. $\phi_n,\phi_c$, and $\phi_s$ represent the node type of news article, subject, and creator respectively.
\end{defn}
We present the schema of News-HIN based on the PolitiFact dataset in Figure~\ref{fig:example}(b), where the exact node and link types can be found intuitively.


\begin{figure}[t]
	\centering
	\begin{minipage}[l]{0.85\columnwidth}
		\centering
		\includegraphics[width=\textwidth]{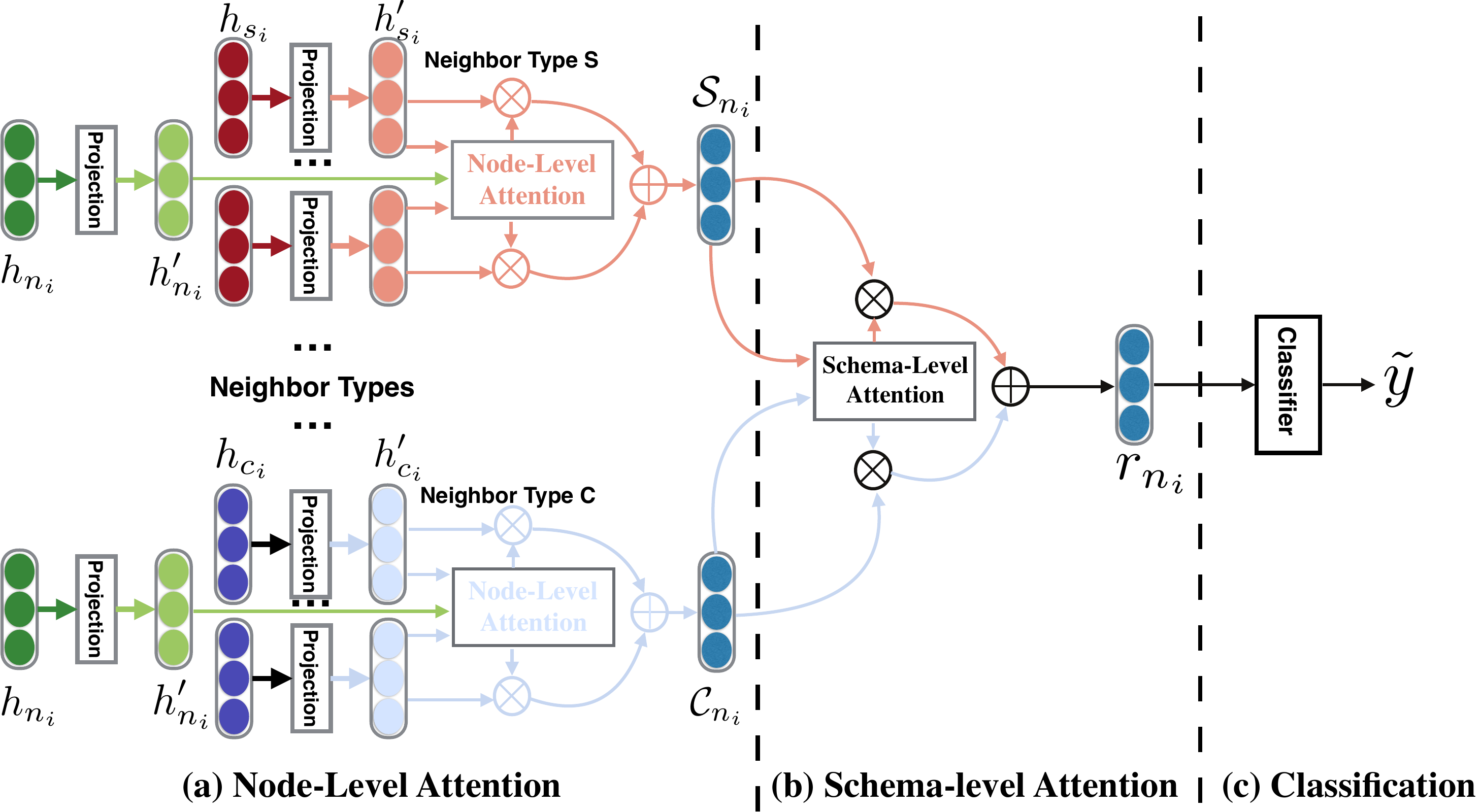}
	\end{minipage}
	\caption{The overall framework of {\our}. (a) All types of nodes are projected into a unified feature space and the weights of node pairs can be learned via node-level attention. (b) Joint learning the weight of each type of schema nodes and fuse representations via schema-level attention. (c) Calculate the loss and end-to-end optimization.}\label{fig:structure}
	\vspace{-15pt}
\end{figure}


\section{Proposed Method}\label{sec:method}

Hierarchical Graph Attention Network ({\our}) follows a hierarchical attention structure including node-level attention and schema-level attention. The structure of {\our} is shown in Figure~\ref{fig:structure}. The node-level attention is proposed to learn the weights of same-typed neighbors and aggregate them to get the type-specific neighbor representation. Then {\our} can learn the information of node types via schema-level attention and achieve the optimal weighted combination for the final fake news detection task. We will further discuss these components in this section. 

\subsection{Node-level attention}\label{subsec:nodeattention}
Node-level attention can learn the importance of neighbors belonging to the same type, respectively, for each news article node $n_i \in \mathcal{N}$. It then aggregates the representations of same-typed neighbors to form an integrated representation that we define as a schema node. 

The inputs of the node-level attention layer are the initial feature vectors of nodes. Because multiple types of nodes exist in the News-HIN, the initial feature vectors belong to feature spaces with different dimensions. In order to enable the attention mechanism to output comparable and meaningful weights, we first utilize a type-specific transformation matrix to project features with different dimensions into the same feature space. We take the news article node $n_i \in \mathcal{N}$ as an example. The transformation matrix for type $\phi_n$ is $\mathbf{M}^{\phi_n} \in \mathbb{R}^{F \times F^{\phi_n}}$, where $F^{\phi_n}$ is the dimension of the initial feature $h_{n_i} \in \mathbb{R}^{F^{\phi_n}}$ and $F$ is the dimension of the feature space mapped to. The $F$ is the same for all type-specific transformation matrices. The projection process can be shown as follows:
\begin{small}
	\begin{equation}
	\begin{aligned}
	h'_{n_i} = \mathbf{M}^{\phi_n} \cdot h_{n_i}; h'_{c_i} = \mathbf{M}^{\phi_c} \cdot h_{c_i}; h'_{s_i} = \mathbf{M}^{\phi_s} \cdot h_{s_i}
	\end{aligned}
	\end{equation}
	
\end{small}

Through the type-specific projection operation, the feature space of nodes with different types can be unified. The node-level attention will learn the importance of same-typed neighbor nodes, respectively. In the face of detecting fake news, the target node is the news article node $n_i \in \mathcal{N}$, and the neighbors of it belong to $\mathcal{N}\cup\mathcal{S}\cup\mathcal{C}$. It should be noted that we also regard the target node itself as a neighbor node to cooperate with the self-attention mechanism. We let $T \in \{\mathcal{N}, \mathcal{S}, \mathcal{C}\}$ and nodes in $T$ have the same type $\phi_t$, then for $n_i$'s neighbor nodes in $T$, the node-level attention can learn the importance $\mathit{e}^{\phi_t}_{ij}$ which means how important node $t_j \in T$ will be for $n_i$. The importance $\mathit{e}^{\phi_t}_{ij}$ can be formulated as follows:
\begin{small}
	\begin{equation}
	\mathit{e}^{\phi_t}_{ij} = \mathit{attention} (h'_{n_i},h'_{t_j};\phi_t)
	\vspace{-15pt}
	\end{equation}
\end{small}

Here, $\mathit{attention}$ denotes the same deep neural network as \cite{VCCRLB18} conducting the node attention. $\mathit{attention}$ is shared for all neighbor nodes with the same type $\phi_t$. The masked attention keeps the network structure information. Only node $t_j \in neighbor_{n_i}$ being neighbors of node $n_i$ with the type $\phi_t$ will be calculated and recorded as $\mathit{e}^{\phi_t}_{ij}$. Otherwise, the attention weight will be 0. We normalize them to get the weight coefficient $\alpha^{\phi_t}_{ij}$ via softmax function:
\begin{small}
	\begin{equation}
	\alpha^{\phi_t}_{ij} = \textup{softmax}_j(\mathit{e}^{\phi_t}_{ij}) = \frac{exp(\mathit{e}^{\phi_t}_{ij})}{\sum_{t_k \in neighbor_{n_i}}\mathit{e}^{\phi_t}_{ik}}
	\end{equation}
\end{small}
The schema node $T_{n_i}$ can be aggregated as follows:
\begin{small}
	\begin{equation}
	T_{n_i} = \sigma(\sum_{t_j \in neighbor_{n_i}}\alpha^{\phi_t}_{ij} \cdot h'_{t_j})
	\vspace{-8pt}
	\end{equation}
\end{small}
\begin{figure}[t]
	\centering
	\begin{minipage}[l]{1\columnwidth}
		\centering
		\includegraphics[width=\textwidth]{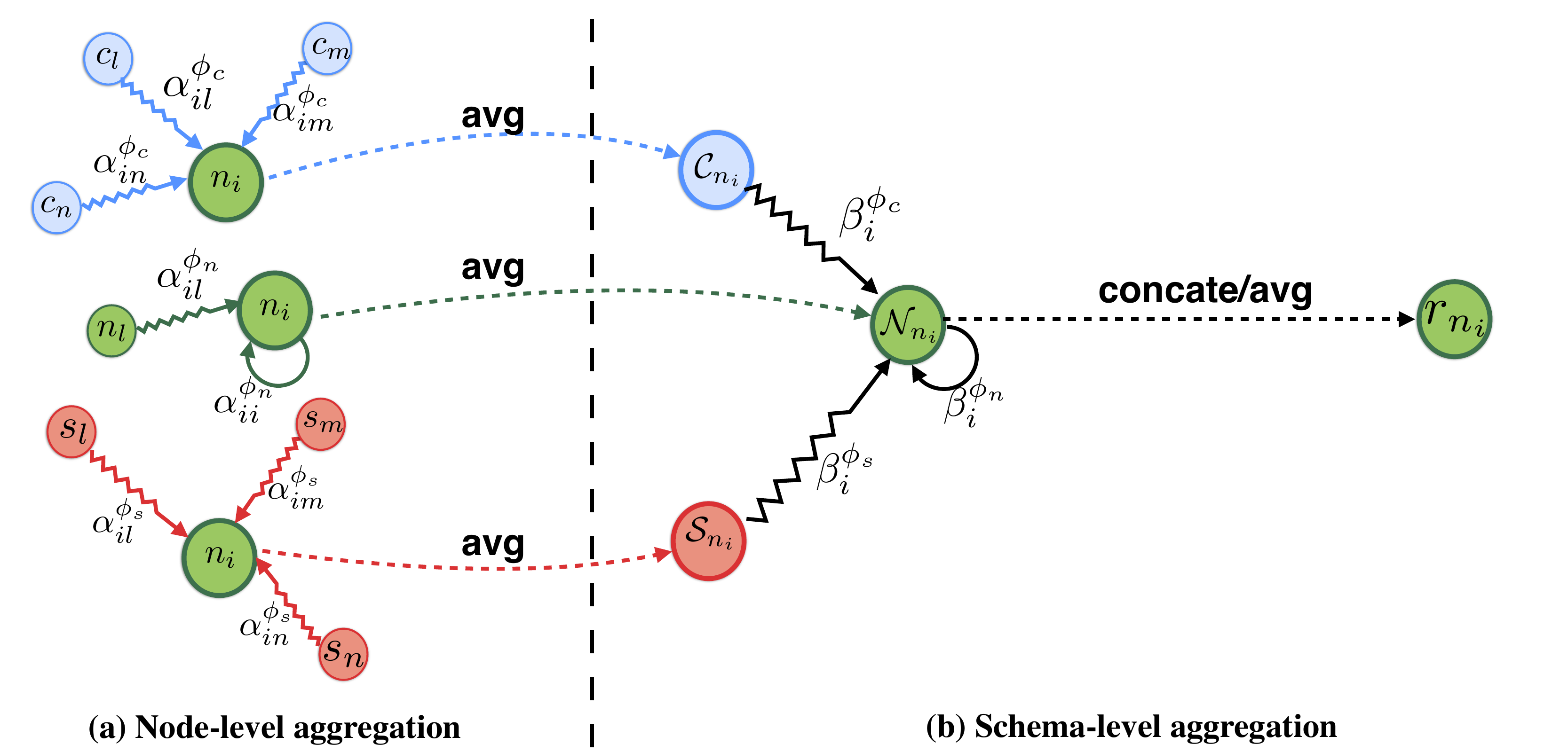}
	\end{minipage}
	\vspace{-10pt}
	\caption{Explanation of aggregating process in both node-level and schema-level.}\label{fig:aggregation}
	\vspace{-18pt}
\end{figure}

A multi-head attention mechanism can be used to stabilize the learning process of self-attention in node-level attention. In details, $K$ independent node-level attentions execute the transformation of Equation (4), and then the features achieved by $K$ heads will be concatenated, resulting in the output representation of the schema node:
\begin{small}
	\begin{equation}
	T_{n_i} = \concatenate_{k=1}^K \sigma(\sum_{t_j \in neighbor_{n_i}}\alpha^{\phi_t}_{ij} \cdot h'_{t_j})
	\vspace{-8pt}
	\end{equation}
\end{small}
where $\concatenate$ represents concatenation. In the problem we face, every target node $n_i$ has 3 schema nodes $\mathcal{N}_{n_i}$, $\mathcal{C}_{n_i}$, $\mathcal{S}_{n_i}$ corresponding to 3 different types neighbors (include itself) based on the Definition~\ref{def:schema}.

\subsection{Schema-level attention}\label{subsec:schemaattention}
Through the node-level attention, we aggregate the neighbors of news article nodes as several schema nodes. Essentially it is equivalent to fusing information from same-typed neighbor nodes into the representation of a schema node. 
What we need to do in this stage is to learn the representation of news article nodes from all schema nodes. Different schema nodes contain type information, which requires us to differentiate the importance of node types. Here we introduce schema-level attention to automatically learn the importance of different schema nodes and use the learned coefficients for weighted fusion.

In order to obtain sufficient expressive power to calculate the attention weights between schema nodes as higher-level features, we apply one learnable linear transformation to the features of schema nodes from node-level attention. The linear transformation is parametrized by a weight matrix $\mathbf{W} \in  \mathbb{R}^{F' \times KF}$. $K$ is the number of heads in node-level attention. The schema-level attention mechanism $\mathit{schema}$ is a single-layer feedforward neural network applying the activating function Sigmoid with the dimension $2F'$. For the schema node $T_{n_i}$, the importance of it can be denoted as $\mathit{w}^{\phi_t}_{i}$:
\begin{small}
	\begin{equation}
	\mathit{w}^{\phi_t}_{i} = \mathit{schema} (\mathbf{W}T_{n_i},\mathbf{W}\mathcal{N}_{n_i})
	\end{equation}
\end{small}
We normalize the importance of each schema nodes through a softmax function. Then the coefficients of the final fusion are denoted as $\beta^{\phi_t}_{i}$, which can be calculated as follows:
\begin{small}
	\begin{equation}
	\beta^{\phi_t}_{i} = \textup{softmax}_{{\phi}_{i}}(\mathit{w}^{\phi_t}_{i}) = \frac{\textup{exp}(\mathit{w}^{\phi_t}_{i})}{\sum_{\phi \in \mathcal{V}_{T}}\textup{exp}(\mathit{w}^{\phi}_{i})}
	\vspace{-5pt}
	\end{equation}
\end{small}
The learned coefficients can indicate the importance of different schema nodes to the final representation. Based on the coefficients, we can aggregate all schema nodes to get the final representation $r_{n_i}$ of the target node $n_i$:
\begin{small}
	\begin{equation}
	r_{n_i} = \sum_{\phi_t \in \mathcal{V}_{T}}\beta^{\phi_t}_{i} \cdot T_{n_i}
	\vspace{-5pt}
	\end{equation}
\end{small}
The set of learned final node representations is denoted as $\mathcal{R}$. Figure~\ref{fig:aggregation} describes the two-level aggregating for reference.

\begin{algorithm}[t]
	
	\caption{{\our}}
		    \scriptsize
	\label{alg:Framwork}
	\KwIn{The News-HIN $\mathcal{G} = (\mathcal{V}, \mathcal{E})$;\\
		\ \ \ \ \ \ \ \ \ \ \ \ \ The initial node feature $h_i, i \in \mathcal{V}, \mathcal{V} = \mathcal{C} \cup \mathcal{N} \cup  \mathcal{S}$\\
		\ \ \ \ \ \ \ \ \ \ \ \ \ The News-HIN Schema $S_{\mathcal{G}} = (\mathcal{V}_{type}, \mathcal{E}_{type})$,\\
		\ \ \ \ \ \ \ \ \ \ \ \ \  $\mathcal{V}_{type} = \{\phi_n,\phi_c,\phi_s\}$\\          
		
	}
	\KwOut{The learned representations $r_{n_i}, {n_i} \in \mathcal{N}$;\\
		\ \ \ \ \ \ \ \ \ \ \ \ \ \ \ The prediction labels vector $\mathcal{L}$}
	\Begin{	
		
		\For{$\phi_t \in \mathcal{V}_{type}$}{
			\For{ nodes $t_i$ of the type $\phi_t$}{
				Feature space projection $h'_{t_i} = \mathbf{M}^{\phi_t} \cdot h_{t_i}$; 
			}		
		}
		\For{$n_i \in \mathcal{N}$}{
			Find the neighbor nodes $neighbor_{n_i}$\;  
			\For{$\phi_t \in \mathcal{V}_{type}$}{
				\For{$t_j \in neighbor_{n_i}$}{
					Calculate the node-level coefficient $\alpha^{\phi_t}_{ij}$; 
				}		
				Aggregate the schema node $T_{n_i} = \sigma(\sum_{t_j \in neighbor_{n_i}}\alpha^{\phi_t}_{ij} \cdot h'_{t_j})$
			}
		} 
		\For{$n_i \in \mathcal{N}$}{
			\For{$\phi_t \in \mathcal{V}_{type}$}{
				Calculate the schema-level coefficient $\beta^{\phi_t}_{i}$; 
			}
			Aggregate to achieve the learned representation $r_{n_i} = \sum_{\phi_t \in \mathcal{V}_{T}}\beta^{\phi_t}_{i} \cdot T_{n_i}$
		}	
		Calculate Cross-Entropy and Back propagation\;
		Update parameters and the prediction labels vector $\mathcal{L}$\;
		\Return{$r_{n_i}, {n_i} \in \mathcal{N}$; $\mathcal{L}$}
	}
	
\end{algorithm}

\subsection{Loss Functions}
Once achieving the final representation, we can use labeled news article nodes to train a classifier. In our experiments, a logistic regression layer is used to make predictions. We define the set of labeled news article nodes as $\mathcal{N}_l$. For the fake news detection tasks, our optimization objective function is set as a cross-entropy loss minimization, and it can be optimized through the backpropagation.

In the binary-class fake news detection, the loss is:
\begin{small}
	\begin{equation}
	\mathit{Loss}(\mathcal{R}, \mathcal{N}_l) = - \sum_{n_i \in \mathcal{N}_l} \big((y_{n_i} \textup{log}(p_{n_i}) + (1-y_{n_i})\textup{log}(1-p_{n_i})\big)
	\label{eq:s_loss}
	\vspace{-5pt}
	\end{equation}
\end{small}
Here, $y$ is a binary indicator (0 or 1) indicating if the label is the correct classification for the news article node. $p_{n_i}$ is the predicted probability of the representation of news article node $n_i$. 

For the multi-class fake news detection, the cross-entropy based loss can be represented as:  
\begin{small}
	\begin{equation}
	\mathit{Loss}(\mathcal{R}, \mathcal{N}_l) = - \sum_{n_i \in \mathcal{N}_l} \sum_{j \in \mathcal{Y}} y_{n_i,j} \textup{log}(p_{n_i,j})
	\label{eq:m_loss}
	\vspace{-5pt}
	\end{equation}
\end{small}
where $y$ is also a binary indicator (0 or 1) indicating whether class label $j$ is the correct classification for the news article node $n_i$. A multi-class logistic regression layer will be trained to output the predicted probability $p_{n_i,j}$ of each class. 

The overall algorithm of {\our} is described in Algorithm~\ref{alg:Framwork}.

\section{Experiments}\label{sec:experiment}
To test the effectiveness of {\our}, extensive experiments are designed and conducted on two real-world fake news datasets. 
Through the experimental results, we plan to evaluate our model by answering the following questions:
\begin{itemize}
	\item \textbf{Question 1}: Can News-HIN and {\our} improve fake news detection performance? 
	\item \textbf{Question 2}: Can the hierarchical attention handle the heterogeneity effectively?
	\item \textbf{Question 3}: Does HGAT possess generalizability so that it can work on different News-HINs and even promote to other applications on heterogeneous graphs?
	
\end{itemize}
\subsection{Dataset Description}
\begin{table}[t]
	\caption{Properties of the Heterogeneous Networks}
	\vspace{-10pt}
	\label{tab:datastat}
	\tiny
	\centering
	\begin{tabular}{cll|ll}
		\toprule	
		&\multicolumn{2}{c}{PolitiFact Network}  &\multicolumn{2}{c}{BuzzFeed Network}\\
		\midrule 
		\multirow{3}{*}{\# node}
		&article   & 14,055 &article   & 182  \\
		&creator  & 3,634 &twitter user   & 15,257 \\
		&subject & 152 &publisher   & 9 \\
		\midrule 
		\multirow{2}{*}{\# link}
		&creator-article    &14,055  &publisher-article    &182   \\
		&article-subject    & 48,756 &article-twitter user    & 25,240 \\
		\bottomrule
	\end{tabular}
	\vspace{-15pt}
\end{table}
One of the datasets used in experiments is collected from the platform with fact-checking: \textit{PolitiFact}, which is operated by Tampa Bay Times. Regarding news articles, \textit{PolitiFact} provides the original contents, fact-checking results, and comprehensive fact-checking reports on the website. The platform categorizes them into different subjects based on contents and topics. A brief description of each subject is provided as well. The fact-checking results can indicate the credibility of corresponding news articles and take values from \{True, Mostly True, Half True, Mostly  False, False, Pants on Fire!\}.
We have established a News-HIN based on the original dataset following the descriptions in the previous sections.
\textit{BuzzFeed}\footnote{https://github.com/KaiDMML/FakeNewsNet/tree/old-version
} from Shu et al.\cite{shu2019beyond}. \textit{BuzzFeed} contains $91$ real news articles and $91$ fake news articles. We also construct a HIN based on \textit{BuzzFeed} dataset following a similar way in Section~\ref{sec:formulation}. There exist three types of nodes: article, twitter user, and publisher. The key statistical data describing two HINs can be found in Table~\ref{tab:datastat}.

\begin{figure*}[t]
	\centering
	\subfigure[\scriptsize Bi-Class Accuracy]{\label{fig:Accuracy}
		\begin{minipage}[l]{1\columnwidth}
			\centering
			\includegraphics[width=1\textwidth]{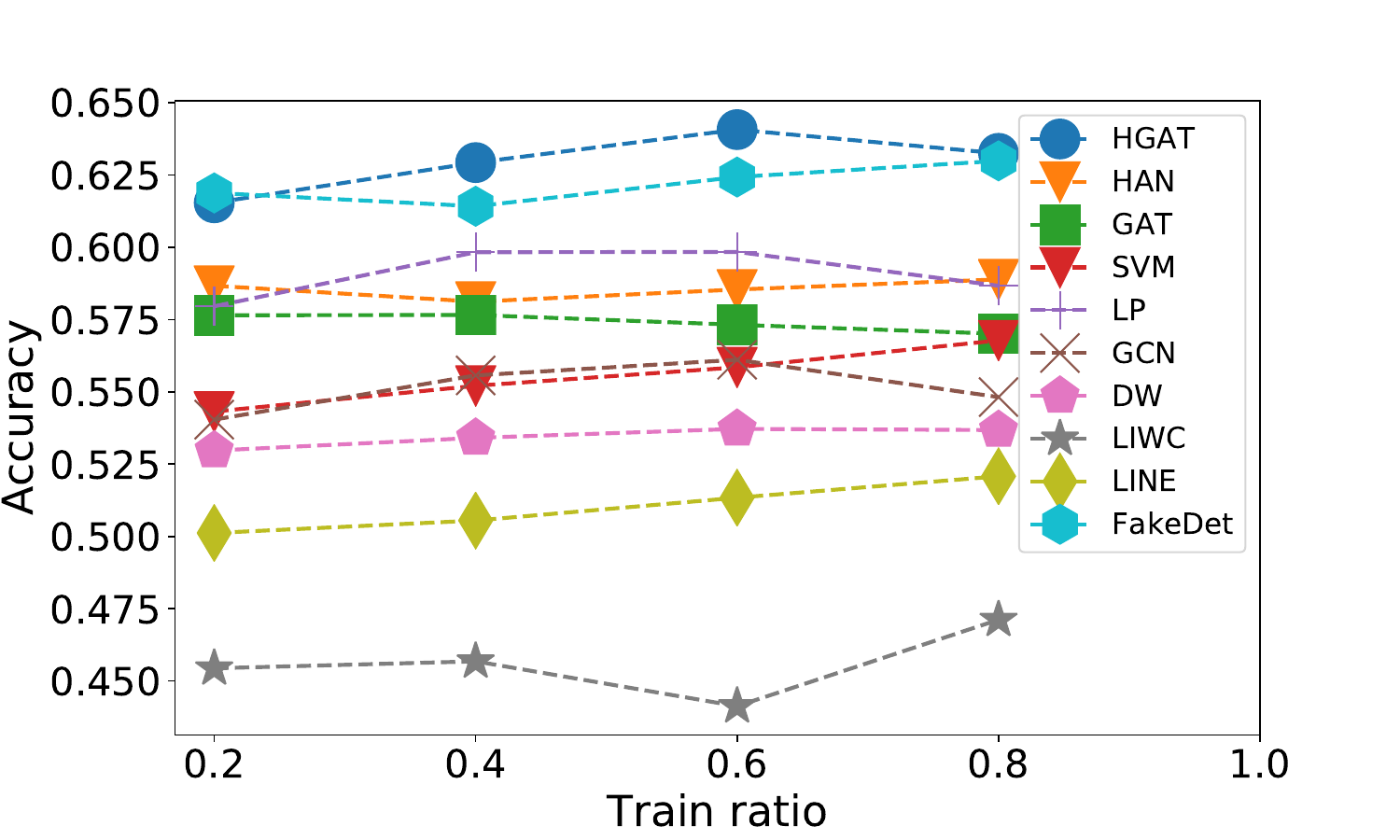}\vspace{-1pt}
		\end{minipage}
	}\hspace{-20pt}
	\subfigure[\scriptsize Bi-Class F1]{\label{fig:F1}
		\begin{minipage}[l]{1\columnwidth}
			\centering
			\includegraphics[width=1\textwidth]{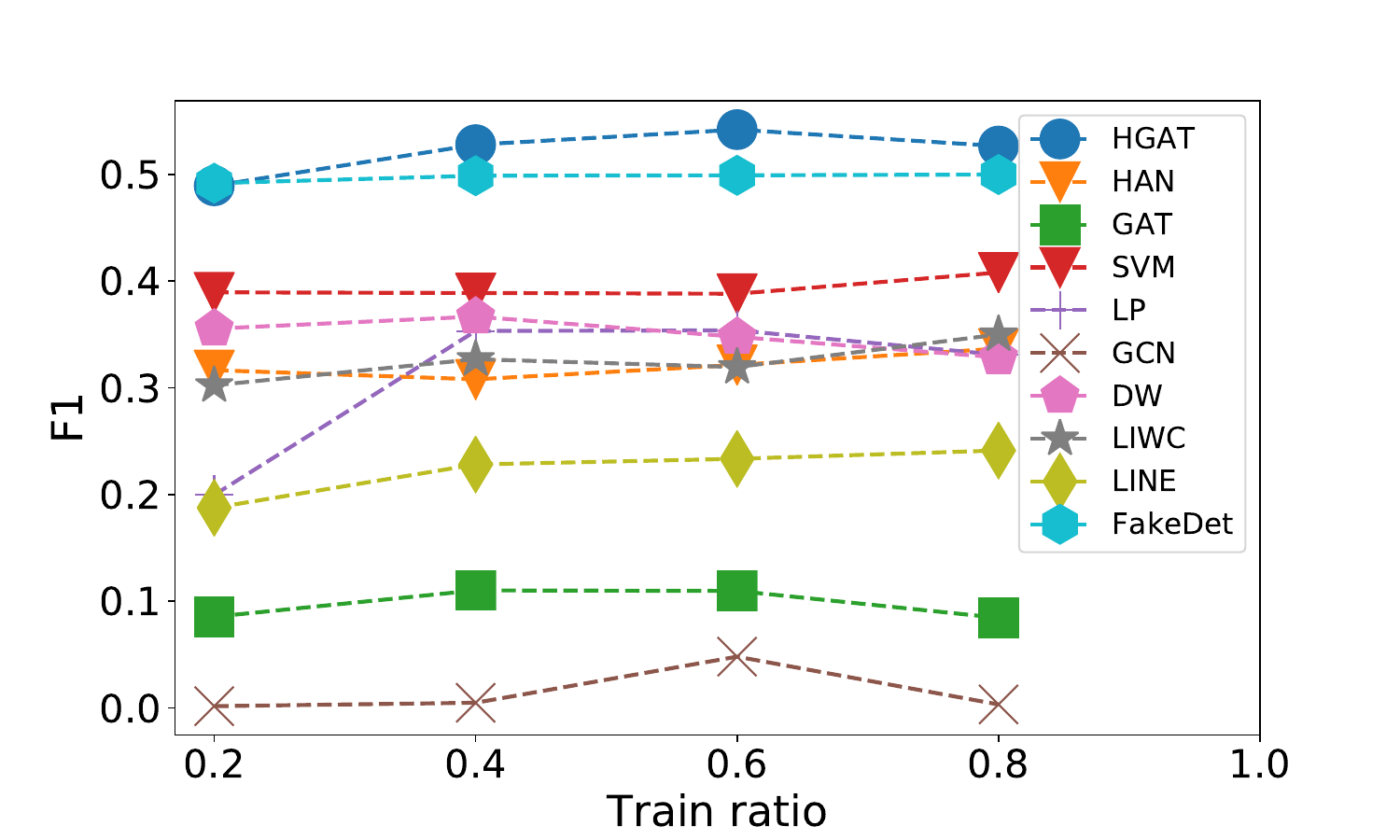}\vspace{-1pt}
		\end{minipage}
	}\vspace{-10pt}
	\subfigure[\scriptsize Bi-Class Recall]{\label{fig:Recall}
		\begin{minipage}[l]{1\columnwidth}
			\centering
			\includegraphics[width=1\textwidth]{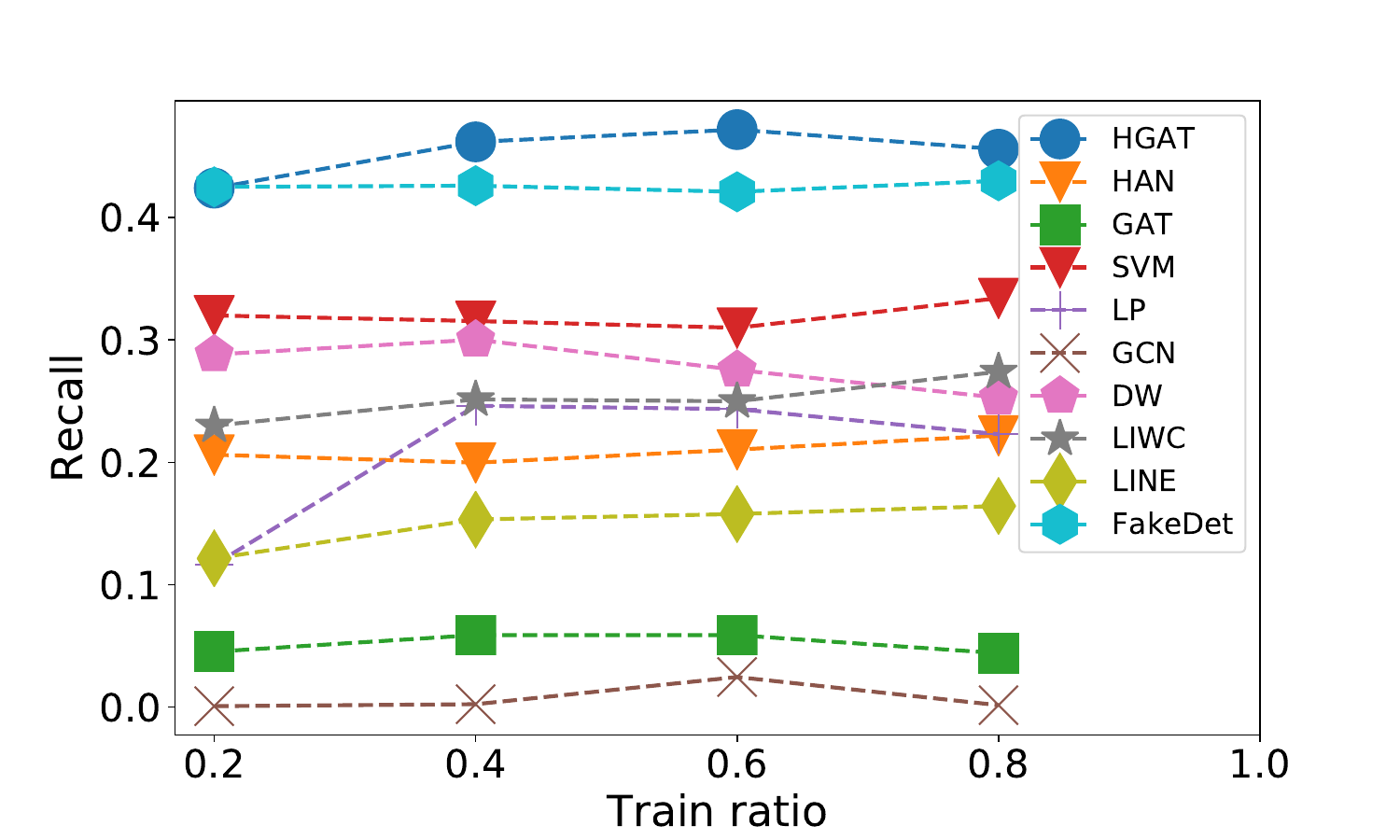}\vspace{-1pt}
		\end{minipage}
	}\hspace{-20pt}
	\subfigure[\scriptsize Bi-Class Precision]{\label{fig:Precision}
		\begin{minipage}[l]{1\columnwidth}
			\centering
			\includegraphics[width=1\textwidth]{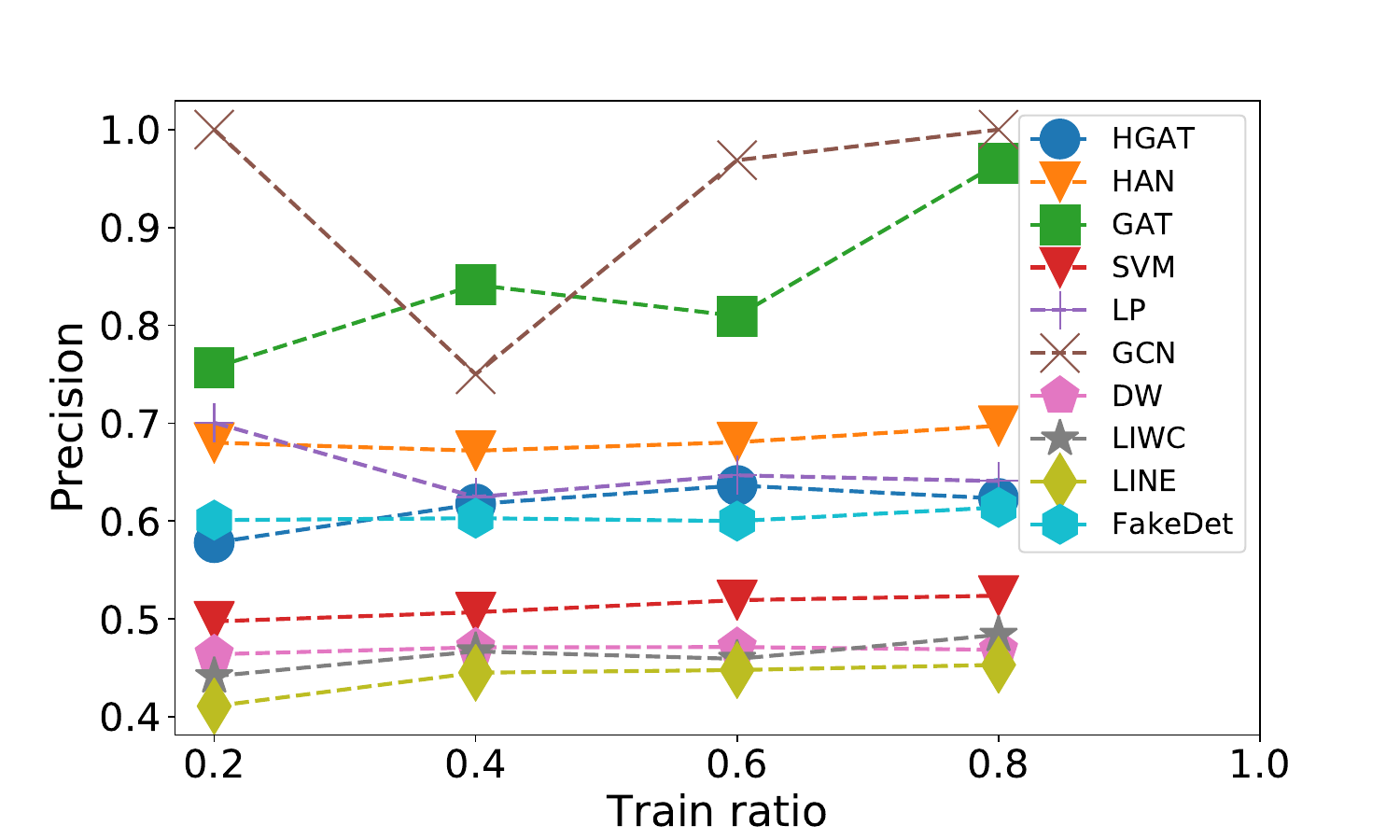}\vspace{-1pt}
		\end{minipage}
	}
	\vspace{-10pt}
	\caption{The results of bi-class news articles classification on \textit{PolitiFact} dataset}\label{fig:bi_class_result}
	\vspace{-20pt}
\end{figure*}
\subsection{Experimental Settings}

\subsubsection{Experimental Setup}
For the \textit{PolitiFact} dataset, we can acquire the set of news article nodes, which are the target nodes, to conduct the classification. The set of news article nodes are divided into 10 folds. Among them, 8 folds will be used as the training set, and 1 fold will be used as the validation set. The remaining 1 fold is left as the testing set. In order to conduct sufficient experiments with the setting of the different numbers of training data, we further make use of 2, 4, 6, 8 of 8 folds as the training set, respectively. In this way, experiments will be conducted with training ratios $\theta \in \{20\%, 40\%, 60\%, 80\%\}$, and the testing ratio is fixed as 10\%. The fact-checking results corresponding to news articles are used as the ground truth for model learning and evaluation. In order to fit the non-sequential models, we have to transform the input features of each type of node
into a vector with a fixed length. We use \textit{TfidfVectorizer} in \textit{Sklearn} package to extract features. The dimensions of news articles, creators, and subjects' initial features are $3000$, $3109$, and $191$, respectively. We will not make use of comprehensive fact-checking reports in our experiments. We train models to work on both the multi-class classification task and the binary-class classification task. In the multi-class classification task, 6 kinds of different fack-checking results correspond to 6 classes. Meanwhile, in the binary-class classification, we group fact-checking results \{Pants on fire, False, Mostly False\} as a Fake class and group \{True, Mostly True, Half True\} as a Real class. Because our target is to detect fake news, we treat the Fake class as the positive class and the Real class as the negative class. In the binary-class classification task, we evaluate the results with Accuracy, Precision, Recall, and F1. Meanwhile, when the model works on the multi-class classification task, the performance is evaluated by Accuracy, Macro Precision, Macro Recall, and Macro F1, respectively. 

For the \textit{BuzzFeed} dataset, it has only two types of labels: True and fake, which can be used directly. In this way, we can only test the binary-class classification task models on \textit{BuzzFeed}. The multi-class classification task will not be conducted on this dataset. The rest settings, including training\&testing sets and initial features extraction, are the same as the \textit{PolitiFact} dataset.


\begin{figure*}[t]
	\centering
	\subfigure[\scriptsize Bi-Class Accuracy]{\label{fig:buzzfeed_Accuracy}
		\begin{minipage}[l]{1\columnwidth}
			\centering
			\includegraphics[width=1\textwidth]{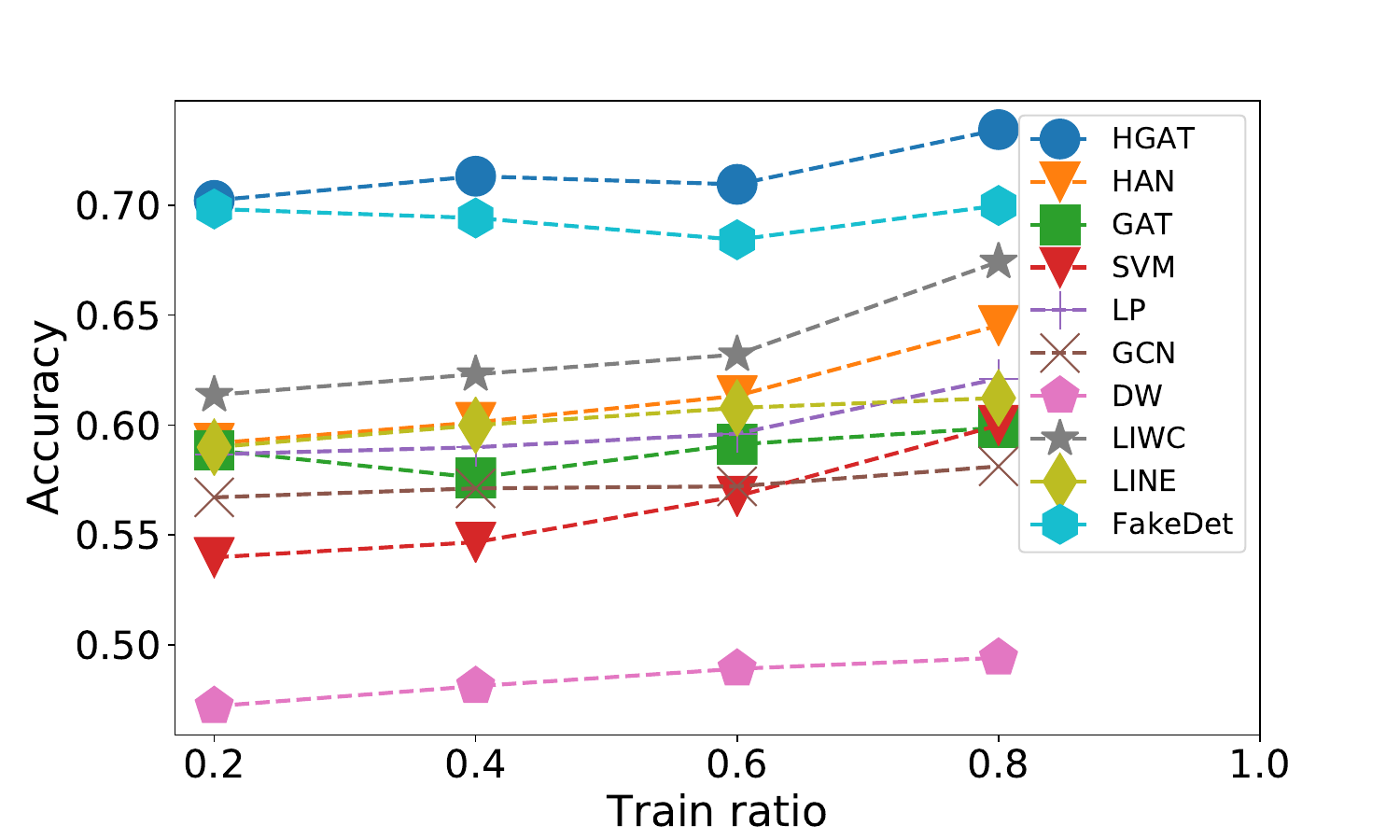}\vspace{-1pt}
		\end{minipage}
	}\hspace{-20pt}
	\subfigure[\scriptsize Bi-Class F1]{\label{fig:buzzfeed_F1}
		\begin{minipage}[l]{1\columnwidth}
			\centering
			\includegraphics[width=1\textwidth]{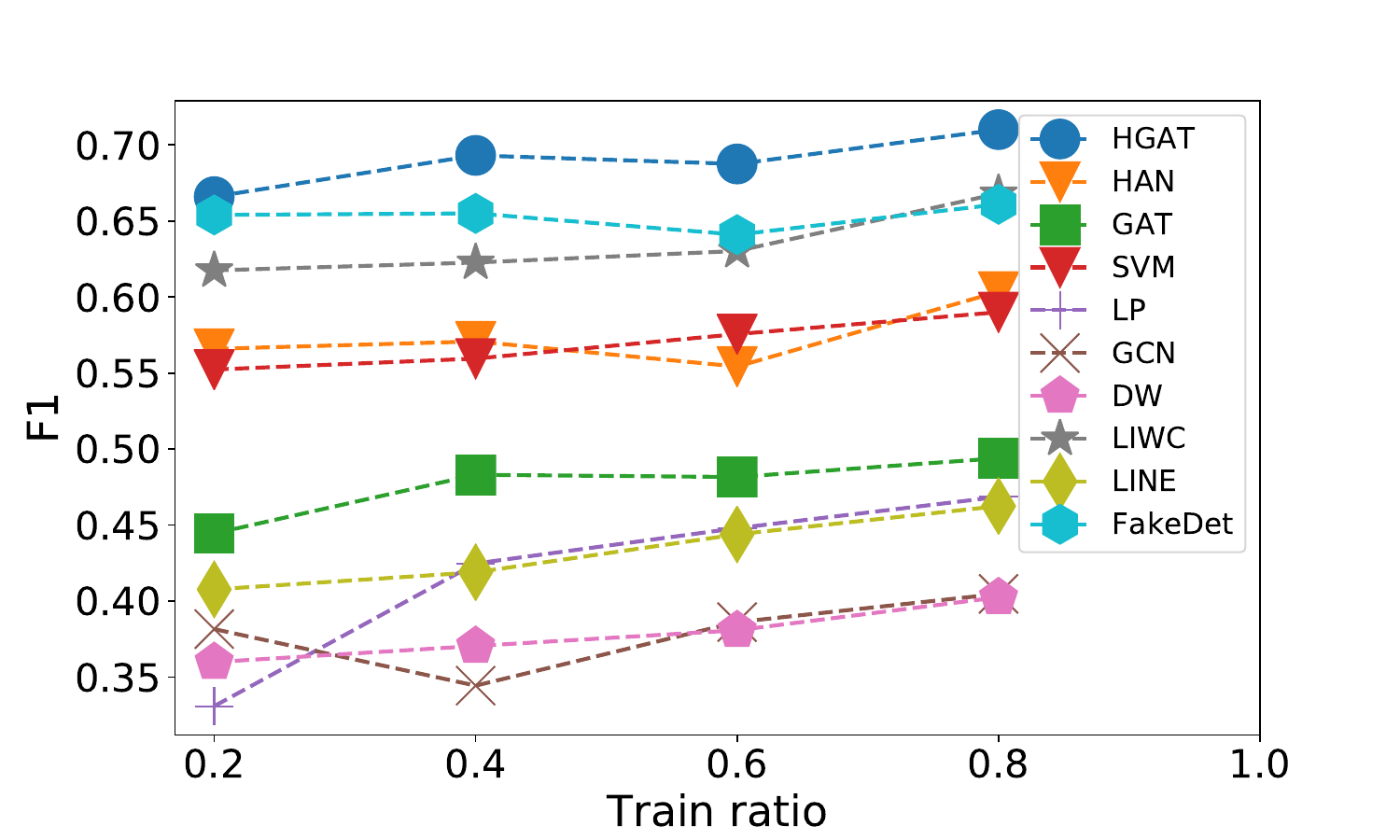}\vspace{-1pt}
		\end{minipage}
	}\vspace{-10pt}
	\subfigure[\scriptsize Bi-Class Recall]{\label{fig:buzzfeed_Recall}
		\begin{minipage}[l]{1\columnwidth}
			\centering
			\includegraphics[width=1\textwidth]{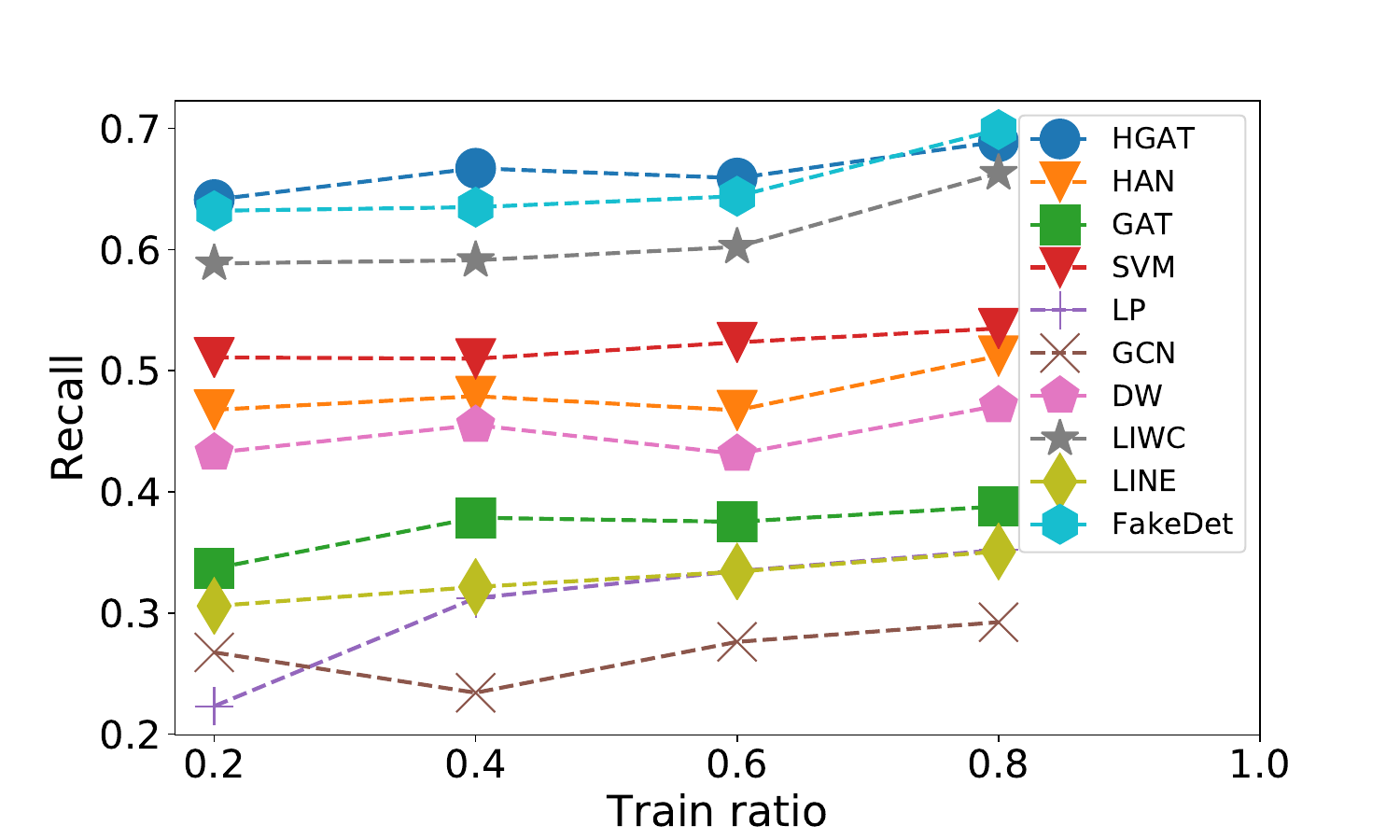}\vspace{-1pt}
		\end{minipage}
	}\hspace{-20pt}
	\subfigure[\scriptsize Bi-Class Precision]{\label{fig:buzzfeed_Precision}
		\begin{minipage}[l]{1\columnwidth}
			\centering
			\includegraphics[width=1\textwidth]{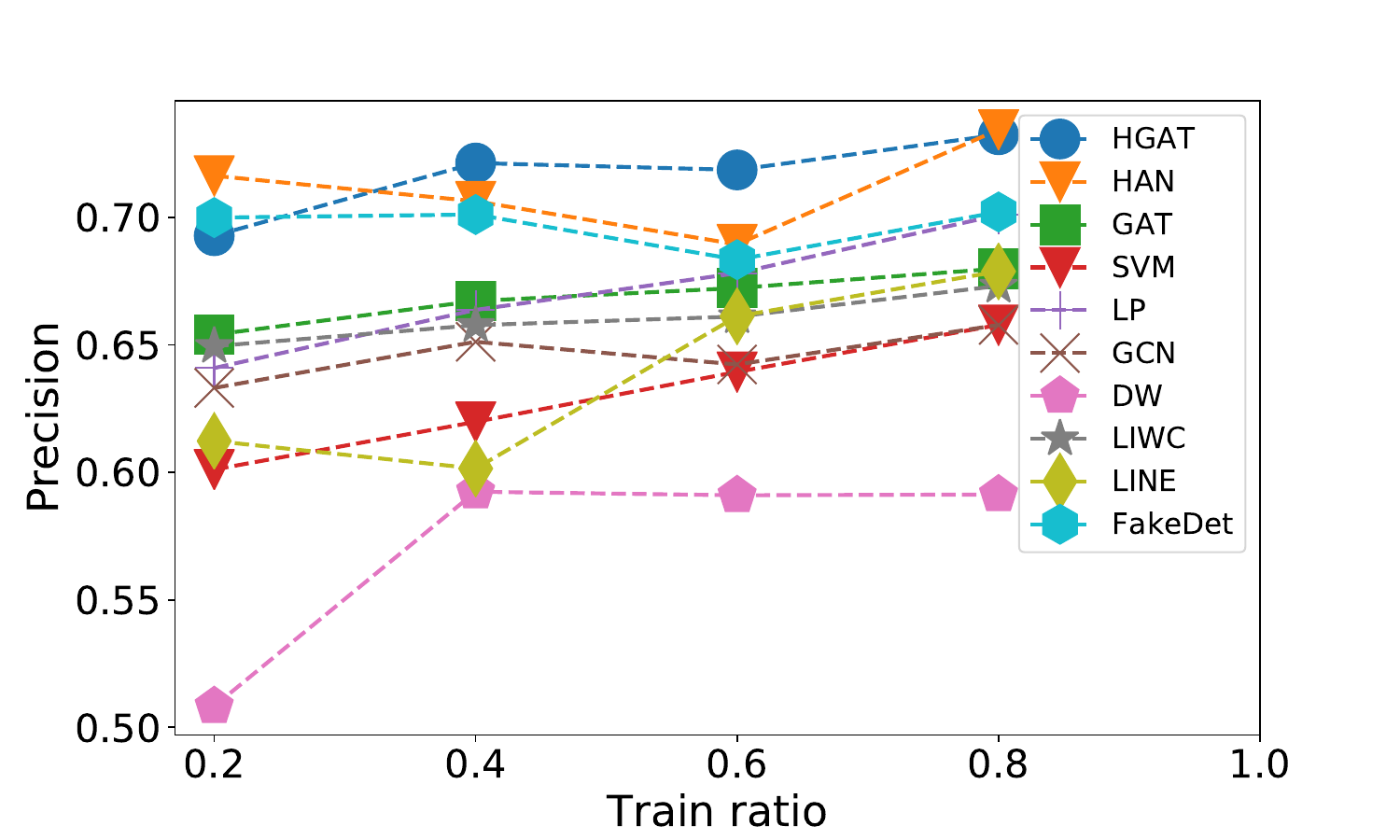}\vspace{-1pt}
		\end{minipage}
	}
	\vspace{-10pt}
	\caption{The results of bi-class news articles classification on \textit{BuzzFeed} dataset}\label{fig:bi_class_result_buzzfeed}
	\vspace{-15pt}
\end{figure*}
\subsubsection{Comparison Methods}
\ \ \ \ \\
\noindent\textbf{\textit{Graph neural network methods}}
\begin{itemize}
	\item \textbf{HAN} \cite{WJSWCYY19}:  HAN employs node-level attention and semantic-level attention to capture the information from all meta-paths. In our experiments, we utilize two meta-paths (article-creator-article, article-subject-article) in HAN.
	\item \textbf{GAT} \cite{VCCRLB18}:  GAT is also an attention-based graph neural network for the node classification, but it is designed for homogeneous graphs. The News-HIN is treated as a homogeneous graph  (ignore the type information) when testing the model.
	\item \textbf{GCN}  \cite{TM17}: GCN is a semi-supervised methods for homogeneous graphs. The News-HIN is treated as a homogeneous graph when testing it.
	\item \textbf{FakeDetector}  \cite{ZCFG18}: FAKEDETECTOR is a deep diffusive network based fake news credibility inference model.
\end{itemize}
\noindent
\textbf{\textit{Text classification methods}}
\begin{itemize}
	\item \textbf{SVM}: SVM is a classic supervised learning model. The feature vector used for building the SVM model is extracted merely based on the news article contents with TF-IDF. 
	\item \textbf{LIWC} \cite{PBJB15}: LIWC stands for Linguistic Inquiry and Word Count, which is widely used to extract the lexicons falling into psycho-linguistic categories. It learns a feature vector from a psychology and deception perspective.
\end{itemize}
\noindent
\textbf{\textit{Network embedding methods}}
\begin{itemize}
	\item \textbf{Label Propagation (LP)} \cite{Zhu02learningfrom}: LP is merely based on the network structure. The prediction scores will be rounded and cast into labels.
	\item \textbf{DeepWalk} \cite{PAS14}: DeepWalk is a random walk based embedding method, which is designed to deal with the homogeneous network. Based on the embedding results, we then train a logistic regression model to perform the classification of news articles.
	\item \textbf{LINE} \cite{JMMMJQ15}: LINE optimizes the objective function that preserves the local and the global network structure simultaneously. We also learn a logistic regression model to conduct the classification based on the learned embeddings.
\end{itemize}
We have also noticed some recently appeared methods for fake news detection \cite{cui2019defend,SWL19}, but did not compare them. The primary consideration is the difference between the scenarios we face. In \cite{cui2019defend,SWL19}, they all utilize social context like user comments, but {\our} aims at detecting fake news in a relatively early stage. We will not utilize user comments about the news because when many users have started to discuss one fake news, the harmful influence of fake news has spread.


\subsection{Reproducibility}
The dimension of node-level representations is set as $12$ and the dimension of schema-level is set as $(8*K)$. Here, the attention head $K$ is set as $1$. For HAN, we set the dimension of node-level representations to $12$ the same as {\our}, and the number of semantic-level hidden units is $8$. 
For GAT, we set the embedding dimension as $12$ and use just $1$ attention head for a fair consideration. For GCN, the embedding dimension is set as $512$. In the DeepWalk, we set the window size to $5$, length of the random walk to $30$, the number of walks per node to $10$, and the embedding dimension to $128$.
We run the experiments on the Server with 3 GTX-1080 ti GPUs, and all codes are implemented in Python3. 
Code is available at:
\href{https://github.com/YuxiangRen/Hierarchical-Graph-Attention-Network}{https://github.com/YuxiangRen/Hierarchical-Graph-Attention-Network.}

\begin{table*}[t]
	\caption{The results of multi-class news articles classification on \textit{PolitiFact} dataset}
	\label{tab:main_result}
	\vspace{-10pt}
	{
		\scriptsize
		\centering
		\begin{tabular}{l||c|cc|ccc|ccccc}
		\cmidrule[2pt]{1-12}
			\multicolumn{2}{c}{}&  \multicolumn{2}{c}{\textbf{Text Classification}}
			& \multicolumn{3}{c}{\textbf{Network Embedding}} &\multicolumn{5}{c}{\textbf{GNNs}} \\
			\cmidrule[1pt]{1-12}
			\textbf{Train}& \textbf{Metric} & SVM & LIWC & LP & DeepWalk & LINE & GCN & GAT & HAN & FakeDetector& {\our} \\
			\midrule
			\multirow{4}{*}{20\%} & Accuracy & 0.1967 & 0.1432 & 0.2218 & 0.1932 & 0.1532 & 0.1986 & 0.2110 & 0.2181 &0.2363 & \textbf{0.2561}\\
			\cmidrule(l){2-12}
			& F1 & 0.1624 & 0.1225 & 0.1925 & 0.1562 & 0.0765 & 0.0711 & 0.1054 & 0.1234&\textbf{0.2204} &0.2141\\
			\cmidrule(l){2-12}
			& Recall & 0.1801 & 0.0965 & 0.2153 & 0.1718 & 0.1433 & 0.1654 & 0.1975 & 0.1884 &0.2307 &\textbf{0.2415}\\
			\cmidrule(l){2-12}
			& Precision & 0.1905 & 0.1409 & 0.2859 & 0.1742 & 0.0326 & 0.0674 & 0.1687 & 0.2467 &\textbf{0.3123} & 0.2397 \\
			\midrule
			\multirow{4}{*}{40\%}& Accuracy & 0.2042 & 0.1543 & 0.2278 & 0.1952 & 0.1567 & 0.1971 & 0.2237 & 0.2240 &0.2399 & \textbf{0.2757}\\
			\cmidrule(l){2-12}
			& F1 & 0.1775 & 0.1314 & 0.1944 & 0.1646 & 0.0798 & 0.0735 & 0.1103 & 0.1441&0.2238 &\textbf{0.2484}\\
			\cmidrule(l){2-12}
			& Recall & 0.1892 & 0.0987 & 0.2183 & 0.1742 & 0.1505 & 0.1697 & 0.1987 & 0.1853 &0.2428 &\textbf{0.2616}\\
			\cmidrule(l){2-12}
			& Precision & 0.2047 & 0.1491 & 0.3037 & 0.1745 & 0.0401 & 0.0685 & 0.1815 & 0.2572 &0.3421 &\textbf{0.3649} \\
			\midrule
			\multirow{4}{*}{60\%} & Accuracy & 0.2061 & 0.1513 & 0.2373 & 0.1969 & 0.1453 & 0.1927 & 0.2214 & 0.2256 &0.2416 &\textbf{0.2707}\\
			\cmidrule(l){2-12}
			& F1 & 0.1871 & 0.1321 & 0.2099 & 0.1647 & 0.0653 & 0.1023 & 0.1162 & 0.1475&0.2215 &\textbf{0.2445}\\
			\cmidrule(l){2-12}
			& Recall & 0.1976 & 0.1002 & 0.2222 & 0.1764 & 0.1410 & 0.1702 & 0.1954 & 0.1852 &0.2401 &\textbf{0.2576}\\
			\cmidrule(l){2-12}
			& Precision & 0.2118 & 0.1561 & 0.2955 & 0.1966 & 0.0307 & 0.1053 & 0.1870 & 0.2792 &0.3311 &\textbf{0.3767} \\
			\midrule
			\multirow{4}{*}{80\%} & Accuracy & 0.2186 & 0.1567 & 0.2407 & 0.2013 & 0.1623 & 0.1955 & 0.2212 & 0.2207 &0.2576 &\textbf{0.2665}\\
			\cmidrule(l){2-12}
			& F1 & 0.1962 & 0.1305 & 0.2187 & 0.1669 & 0.0875 & 0.1029 & 0.1037 & 0.1218&0.2221 &\textbf{0.2393}\\
			\cmidrule(l){2-12}
			& Recall & 0.2081 & 0.0954 & 0.2341 & 0.1830 & 0.1512 & 0.1940 & 0.1975 & 0.1840 &0.2544 &\textbf{0.2586}\\
			\cmidrule(l){2-12}
			& Precision & 0.2233 & 0.1553 & 0.3149 & 0.1896 & 0.0468 & 0.0977 & 0.1819 & 0.2497 &0.3510 &\textbf{0.3757} \\
		    \cmidrule[2pt]{1-12}
		\end{tabular}
		
		\label{tab:classification_result}
	}
	
\end{table*}
\subsection{Experimental Results with Analysis}

\subsubsection{Evaluate the effectiveness of News-HIN}
To answer \textbf{Question 1}, we can analyze the experimental result from both binary-class and multi-class classification tasks. Based on results in Figure~\ref{fig:bi_class_result}, our model {\our} achieves the best performance when focusing on Accuracy, F1, and Recall. Here we need to point out the reason for the inconsistency between the experimental results of FakeDetector shown in Figure~\ref{fig:bi_class_result} and the reported results in \cite{ZCFG18}. Because the fake news is set to positive class in our experiments, while the real news is set to positive in \cite{ZCFG18}, the different settings lead to inconsistency. Our main task is to detect fake news, so we set fake news as positive to make the experimental results more intuitive. When considering precision, we can observe from Figure~\ref{fig:Precision} that the performance of {\our} is lower than that of GCN and GAT. Through careful analysis, it can be found that GAT and GCN tend to judge most instances as 'Real' in the face of fake news detection, so the higher precision is related to very low Recall. In this case, higher precision is not practical because much fake news can not be detected. By comparing the performance between {\our} and network embedding methods, we can conclude that textual information is quite important, and merely utilizing graph structure is insufficient. Simultaneously, through the comparison between {\our} and text classification methods, we can find that graph structure is powerful for fake news detection. As a data structure that can model the graph structure and textual information simultaneously, news-HIN achieves better results. {\our} also shows an advantage over HAN, which is also proposed for heterogeneous graphs. More important. {\our} is a meta-path-free model without the limitation of handcrafted features. Figure~\ref{fig:bi_class_result_buzzfeed}, showing the results from the other dataset, also validates our conclusions.
\begin{figure*}[t]
	\centering
	\subfigure[\scriptsize Accuracy]{\label{fig:two_Accuracy}
		\begin{minipage}[l]{0.5\columnwidth}
			\centering
			\includegraphics[width=1\textwidth]{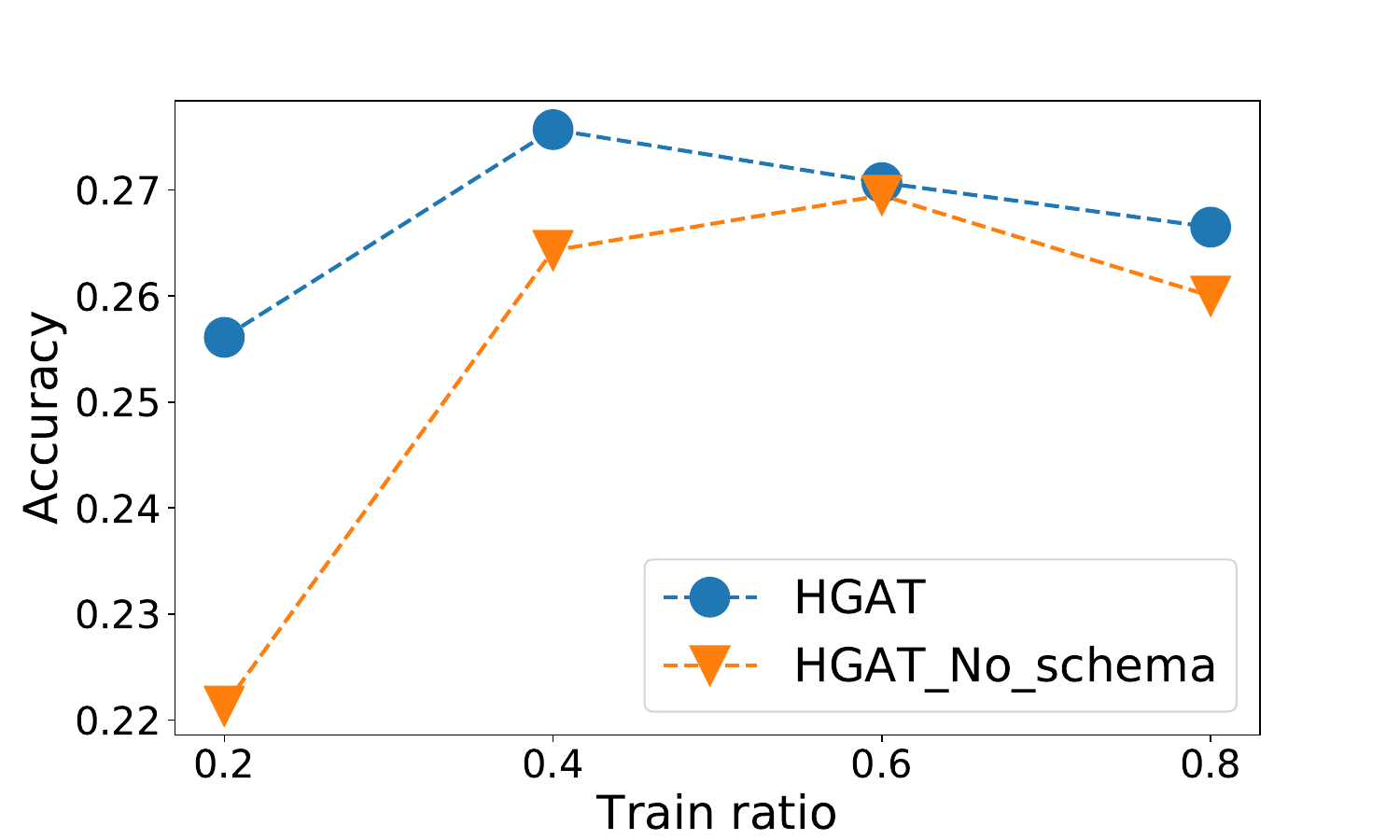}\vspace{1pt}
		\end{minipage}
	}\hspace{-.15in}
	\subfigure[\scriptsize Macro-F1]{\label{fig:two_F1}
		\begin{minipage}[l]{0.5\columnwidth}
			\centering
			\includegraphics[width=1\textwidth]{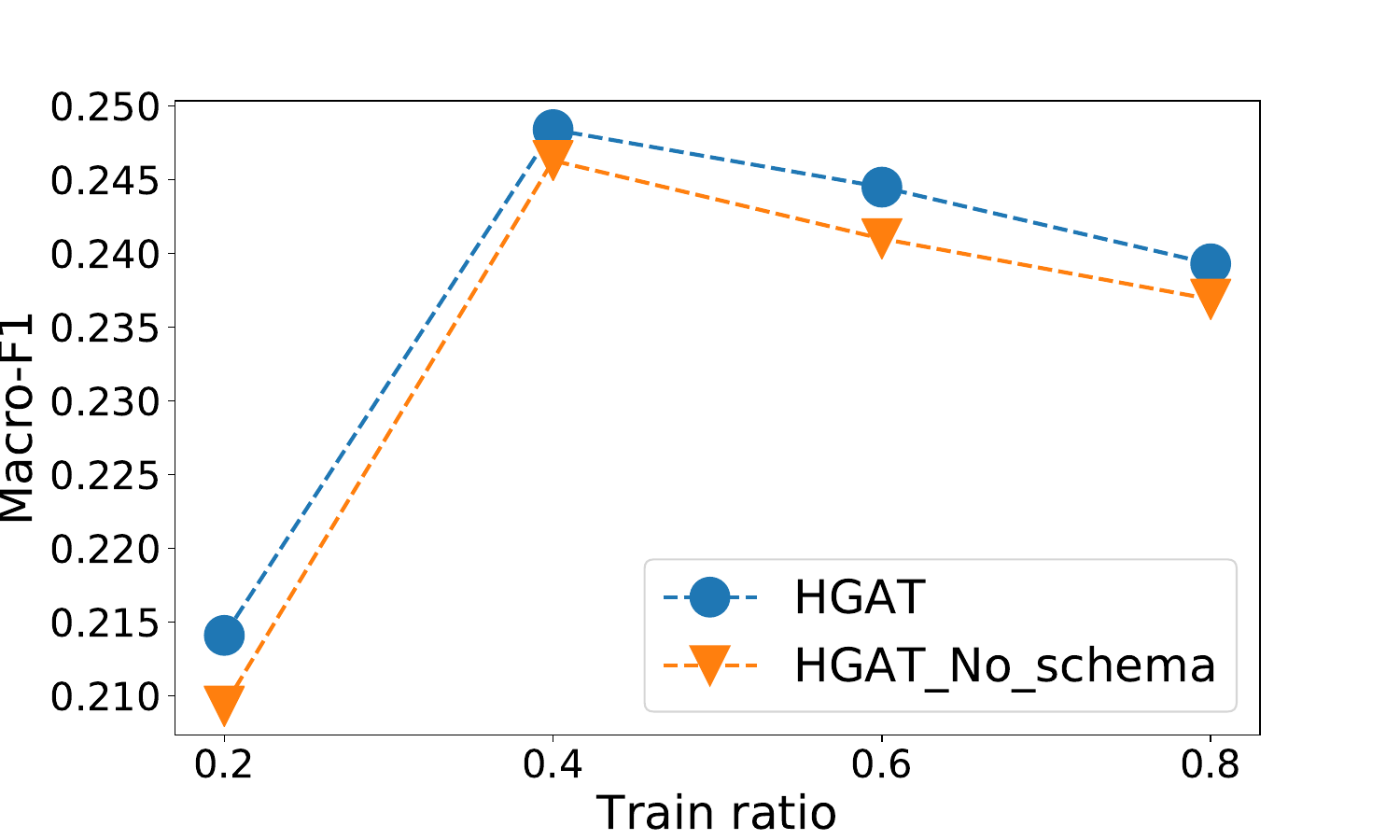}\vspace{1pt}
		\end{minipage}
	}\hspace{-.15in}
	\subfigure[\scriptsize Macro-Recall]{\label{fig:two_Recall}
		\begin{minipage}[l]{0.5\columnwidth}
			\centering
			\includegraphics[width=1\textwidth]{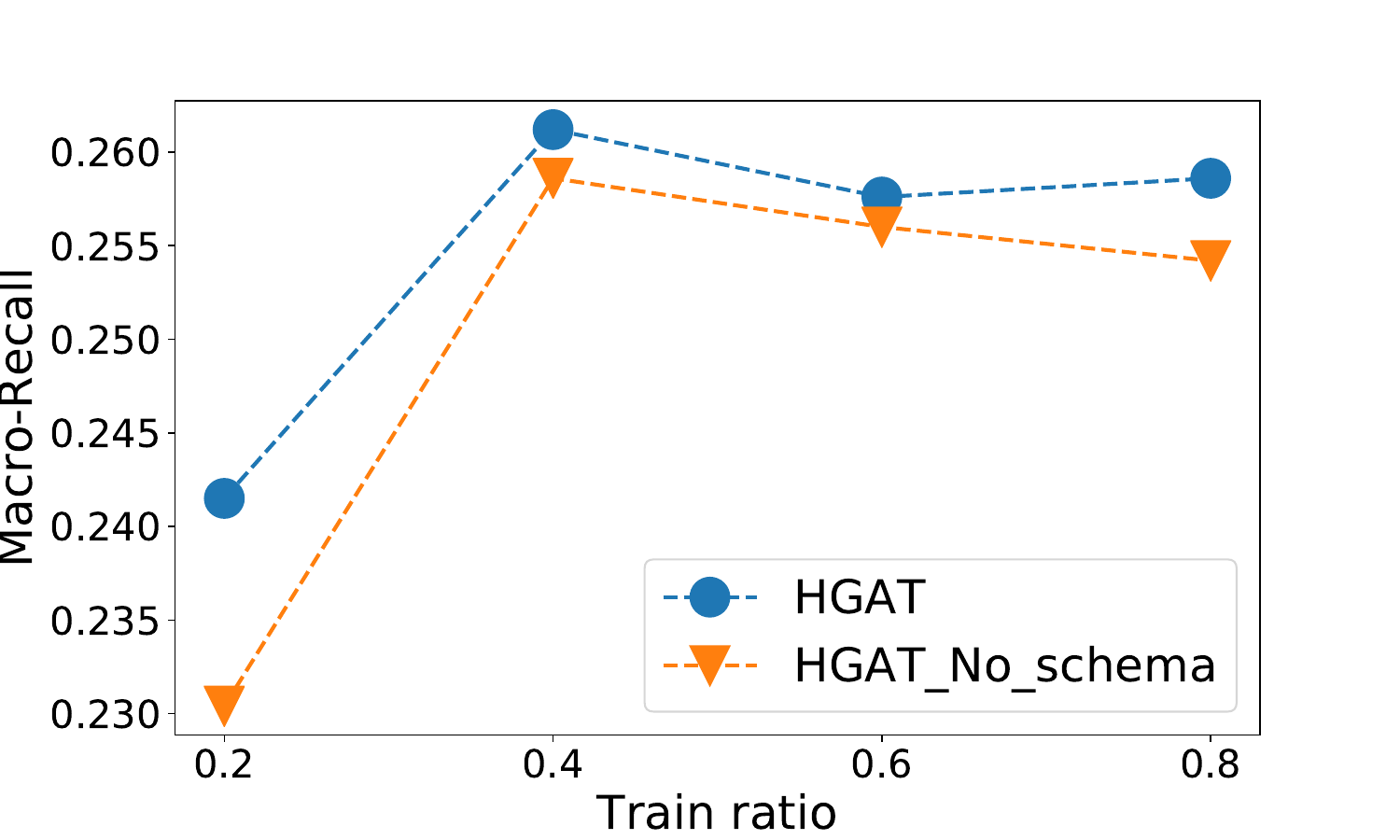}\vspace{1pt}
		\end{minipage}
	}\hspace{-.15in}
	\subfigure[\scriptsize Macro-Precision]{\label{fig:two_Precision}
		\begin{minipage}[l]{0.5\columnwidth}
			\centering
			\includegraphics[width=1\textwidth]{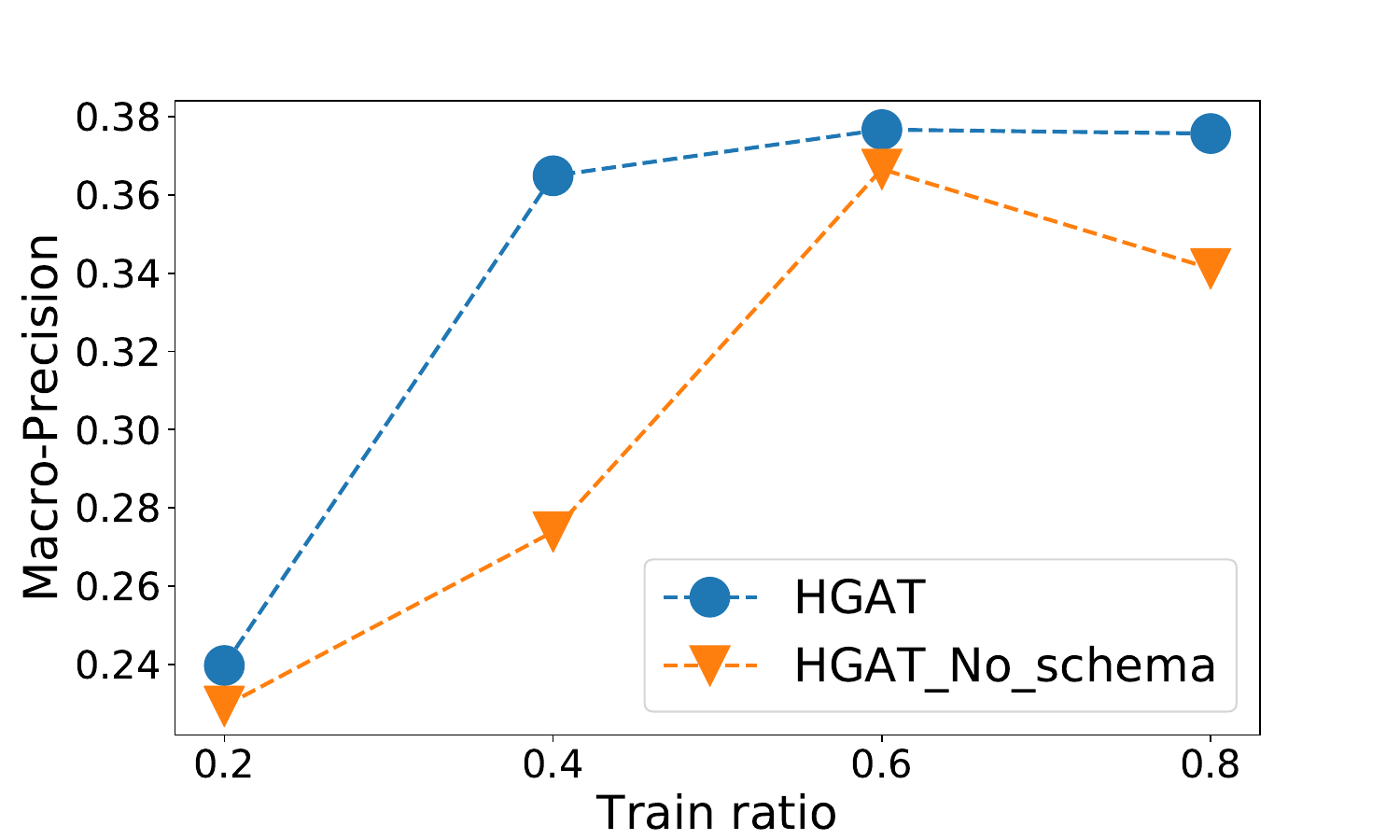}\vspace{1pt}
		\end{minipage}
	}\hspace{-.1in}
	\vspace{-10pt}
	\caption{The comparison of multi-class news article classification between {\our} and {\our} without schema-level attention on \textit{PolitiFact} dataset}\label{fig:comparison_result}
	\vspace{-18pt}
\end{figure*}
Next, we continue to analyze the multi-class classification task results to answer \textbf{Question 1} further. Due to the uncertainty of emerging news, it is often difficult to judge news directly as absolutely true and false. Besides, it is also not conducive to subsequent operations (e.g., final verification). Carrying out finer granularity multi-classification tasks according to the credibility of news is very meaningful. The experimental results from 6-labels classification are shown in Table~\ref{tab:main_result}. Analyzing from the results, {\our} also achieves satisfactory performance. Compared with text classification methods and network embedding methods, {\our} outperforms them with an obvious advantage. On the one hand, this set of results illustrates the significance of News-HIN. On the other hand, this also shows that {\our} has a stronger learning ability in the heterogeneous graph, and the learned representation is also more comprehensive and discriminative. As we utilize a News-HIN as source data, the heterogeneity should be handle in an effective manner. We will evaluate the performance of {\our} in handling heterogeneity detailedly in the next section.

\subsubsection{Access the performance in handling heterogeneity}
In order to answer \textbf{Question 2}, we further compare the performance of GNN methods. From Figure~\ref{fig:bi_class_result} and Figure~\ref{fig:bi_class_result_buzzfeed} to Table~\ref{tab:main_result}, different tasks on two datasets verify that the heterogeneity of graphs should be dealt with in a more effective way. If we treat a heterogeneous graph as a homogeneous graph by ignoring the type (as we did in GCN and GAT), the result would be very disappointing. Compared with the method HAN proposed for heterogeneous graphs, {\our} also shows an advantage. 

To further demonstrate the effectiveness of schema-level attention, we replace the schema-level attention of {\our} with fixed and equal weights for schema nodes. In experiments, all three schema nodes are assigned with the weight $1/3$, and we denote this comparison model as HGAT\_No\_schema. In Figure~\ref{fig:comparison_result}, we present the comparison results between {\our} and HGAT\_No\_schema. It is obvious that {\our} achieves better performance than HGAT\_No\_schema according to various metrics on two different datasets. This comparison verifies that the importance of schema nodes worth distinguishing, and {\our} can differentiate the importance through attention weights effectively. In contrast, the simple average operation harms the performance, which is equivalent to dropping the type information of schema nodes.
%

\subsubsection{Verify the generalizability of HGAT}
To answer \textbf{Question 3}, we design experiments for {\our} to test performance when facing different News-HINs. \textit{PolitiFact} and \textit{BuzzFeed} provide two different News-HINs. 
What's more, heterogeneous graph attention network~\cite{WJSWCYY19} (HAN), as a general representation learning method proposed for heterogeneous graphs, is limited to handcrafted features (i.e., meta-path). {\our} can beat HAN without the limitation of any manual features, which fully illustrates the generalizability of {\our}. When facing other node classification-related applications on heterogeneous graphs, {\our} can be utilized as a general node representation learning method and be transferred without any obstacle.
Although we only focus on fake news detection in this paper, {\our} is not restricted by the graph structure and is essentially a general method. Other potential applications of {\our} will be left for future discussion.

\section{Conclusion}\label{sec:conclusion}
In this paper, we study the HIN-based fake news detection problem and propose a novel graph neural network {\our} to solve it. {\our} employs a hierarchical attention mechanism considering both node-level and schema-level attention to learn the comprehensive representations of news article nodes. These discriminative representations can be used to detect fake news. Extensive experiments on two real-world News-HIN demonstrate that {\our} has outstanding performance compared with the state-of-the-art methods. {\our} is also a general graph representation learning model that does not require any handcrafted features (e.g. meta-path) or other prior knowledge, so it is highly extensible for node classification-related problems on heterogeneous graph other than fake news detection. 

For future work, we have some interesting plans to share with you. At first, in this paper, a relatively simple method (i.e., TF-IDF) is used to process the textual information of the node. In the future, it may be considered to integrate more powerful methods in natural language processing into heterogeneous graphs to construct a novel end-to-end model. Secondly, fact-checking reports are worth exploring, which can help the model learn from experts to detect fake news. At last, it is meaningful and interesting to support the fake news detection by mining the behavior patterns of high-risk fake news publishers.

\balance
\bibliographystyle{plain}
\bibliography{refs}

\end{document}